# Towards Multi-Scale Modeling of Carbon Nanotube Transistors

Jing Guo, Supriyo Datta, and Mark Lundstrom
School of Electrical and Computer Engineering, Purdue University
West Lafayette, IN 47907

M. P. Anantam
NASA Ames Research Center, Moffett Field, CA 94035

## *Abstract*

Multiscale simulation approaches are needed in order to address scientific and technological questions in the rapidly developing field of carbon nanotube electronics. In this paper, we describe an effort underway to develop a comprehensive capability for multiscale simulation of carbon nanotube electronics. We focus in this paper on one element of that hierarchy, the simulation of ballistic CNTFETs by self-consistently solving the Poisson and Schrödinger equations using the non-equilibrium Green's function (NEGF) formalism. The NEGF transport equation is solved at two levels: i) a semi-empirical atomistic level using the $p_z$ orbitals of carbon atoms as the basis, and ii) an atomistic mode space approach, which only treats a few subbands in the tube's circumferential direction while retaining an atomistic grid along the carrier transport direction. Simulation examples show that these approaches describe quantum transport effects in nanotube transistors. The paper concludes with a brief discussion of how these semi-empirical device level simulations can be connected to *ab initio*, continuum, and circuit level simulations in the multi-scale hierarchy.

## 1. Introduction

Carbon nanotubes show promise for applications in future electronic systems, and the performance of carbon nanotube transistors, in particular, has been rapidly advancing [Win02, Jav03]. From a scientific perspective, carbon nanotube electronics offers a model system in which to explore and understand the effects of detailed microstructure of contacts, interfaces, and defects. It is also an opportunity to develop the theory and computational techniques for the atomistic simulation of small electronic devices in general. A detailed treatment of carbon nanotube electronics requires an atomistic description of the nanotube along with a quantum mechanical treatment of electron transport, both ballistic and with the effects of dissipative scattering included. As shown in Fig. 1, even for this simple system, multi-scale methods are essential. Metal/nanotube contacts, nanotube/dielectric interfaces, and defects require a rigorous, *ab initio* treatment, but to treat an entire device, simpler, $p_z$ orbital descriptions must be used. Techniques connect different descriptions used for different regions of the device will need to be developed (e.g. the ab initio basis functions for the metal/nanotube contacts must be connected to the semi-empirical basis functions for the device itself). For extensive device optimization, continuum, effective mass level models may be necessary, and methods to relate the phenomenological parameters in those approaches to the atomistic models must be developed. For circuit simulation, even simpler, analytical models are needed, and efficient techniques for extracting circuit models from physically detailed models must be devised.

Our purpose in this paper is to describe the status of our work to develop a comprehensive, multi-scale simulation capability for electronic devices. We will focus on our initial effort that make use of a semi-empirical, $p_z$ orbital description, and discuss briefly the challenges to be addressed in connecting this work to *ab initio* simulations, to continuum device simulations, and to circuit models. The approach has already demonstrated its usefulness in analyzing recent experimental data, suggesting experiments, and in exploring device possibilities [Guo03a, Guo03b].

## 2. Review of the NEGF Formalism

A carbon nanotube can be viewed as a rolled-up sheet of graphene with a diameter typically between one and two nanometers. The nanotube can be either metallic or semiconducting, depending on how it is rolled up from the graphene sheet (i.e. depending on its chirality) [Sai98]. Semiconducting nanotubes are suitable for transistors. In order to correctly treat carbon nanotube transistors, strong quantum confinement around the tube circumferential direction, quantum tunneling through Schottky barriers at the metal/nanotube contacts, and quantum tunneling and reflection at barriers in nanotube channel need to be considered. The non-equilibrium Green's function (NEGF) formalism, which solves Schrödinger equation under non-equilibrium conditions and can treat coupling to contacts and dissipative scattering process, provides a sound basis for quantum device simulations. **[**Dat95**]**. The NEGF simulation approach has demonstrated its usefulness for simulating nanoscale transistors from conventional Si MOSFETs, [Ren03], MOSFETs with novel channel materials [Rah03], to CNTFETs [Guo03a], and molecular transistors [Dam02]. In this section, we give brief summary of the NEGF simulation procedure. For a more thorough description of the technique, see [Dat00, Dat02].

Figure 2 shows a generic transistor and defines some terms for the NEGF simulation. The first step is to identify a suitable basis set and Hamiltonian matrix for an isolated channel. The self-consistent potential, which is a part of the Hamiltonian matrix, is included in this step. The second step is to compute the self-energy matrices, $\Sigma_1$, $\Sigma_2$ and $\Sigma_S$, which describe how the ballistic channel couples to the source/drain contacts and to the scattering process. (For simplicity, only ballistic transport is treated in this paper.) After identifying the Hamiltonian matrix and the self-energies, the third step is to compute the retarded Green's function,

$$\mathbf{G}(E) = [(E + i0^+)\mathbf{I} - \mathbf{H} - \mathbf{\Sigma}_1 - \mathbf{\Sigma}_2]^{-1}. \tag{1}$$

The fourth step is to determine the physical quantities of interest from the Green's function.

In the ballistic limit, states within the device can be divided into two parts: 1) states filled by carriers from the source according to the source Fermi level, and 2) states filled by the drain according to the drain Fermi level. Within the device, the source (drain) local-density-of-states (LDOS) is $\mathbf{D}_{S(D)} = \mathbf{G}\mathbf{\Gamma}_{S(D)}\mathbf{G}^{+}$, where $\mathbf{\Gamma}_{S(D)} = i(\mathbf{\Sigma}_{1(2)} - \mathbf{\Sigma}_{1(2)}^{+})$ is the energy level broadening due to the source (drain) contact. The charge density within the device is computed by integrating the LDOS, weighted by the appropriate Fermi level) over energy. The charge contributed by the source contact is

$$Q_S(z) = (-e)\int_{E_N}^{+\infty} D_S(E,z) f(E - E_{FS})dE + e\int_{-\infty}^{E_N} D_S(E,z)\{1 - f(E - E_{FS})\}dE$$

where e is the electronic charge, and $E_N$ is the charge neutrality level [Ter84]. The total charge is

$$Q(z) = Q_S(z) + Q_D(z) = (-e)\int_{-\infty}^{+\infty} dE \cdot \mathrm{sgn}[E - E_N(z)]\{D_S(E,z) f(\mathrm{sgn}[E - E_N(z)](E - E_{FS})) + D_D(E,z) f(\mathrm{sgn}[E - E_N(z)](E - E_{FD}))\}, \quad (2)$$

where $\mathrm{sgn}(E)$ is the sign function, and $E_{FS,D}$ is the source (drain) Fermi level. For a self-consistent solution, the NEGF transport equation is solved with iteratively the Poisson equation until self-consistency is achieved after which the source-drain current is computed from

$$I = \frac{4e}{h}\int T(E)[f_S(E) - f_D(E)]dE \quad (3)$$

where $T(E) = \mathrm{Trace}(\mathbf{\Gamma}_1\mathbf{G}\mathbf{\Gamma}_2\mathbf{G}^{+})$ is the source/drain transmission and the extra factor of two comes from the valley degeneracy in the carbon nanotube energy band structure.

The computationally expensive part of the NEGF simulation is finding the retarded Green's function, according to eqn. (1), which requires the inversion of a matrix for each energy grid point. The straightforward way is to explicitly invert the matrix, whose size is the size of the basis set. This, however, is impractical for an atomistic simulation of a nanotube transistor. In the ballistic limit, the problem is simplified because only a few columns of the Greens's function are needed.  Still, reducing the size of the Hamiltonian matrix and developing computationally efficient approaches are of great importance for an atomistic simulation.

## 3. Atomistic NEGF Treatment of Electron Transport in Carbon Nanotubes

### 3.1 Real Space Approach

In this section, we describe an NEGF simulation of ballistic CNTFETs using a real space basis. The first step is to identify a set of atomistic orbitals adequate to describe the essential physics for carrier transport and then to write down the Hamiltonian matrix for the isolated channel in that basis.  An (n, 0) zigzag nanotube as shown in Fig. 3 is assumed, but the method can be readily extended to armchair or chiral nanotubes.  There are four orbitals in the outer electron shell of a carbon atom (s, $p_x$, $p_y$, and $p_z$).  One $p_z$ orbital is often sufficient because the bands involving $p_z$ orbitals are largely uncoupled from the bands involving the other orbitals, and the bands due to the s, $p_x$ and $p_y$ orbitals are either well below or well above the Fermi level and, therefore, unimportant for carrier transport. With one $p_z$ orbital per carbon atom as the basis set, the size of the Hamiltonian matrix is the number of carbon atoms in the transistor channel.  For typical problems, such as the examples in Sec. 5, a carbon nanotube transistor will consist of several thousand carbon atoms.  We use a tight-binding approximation to describe the interaction between carbon atoms, and only nearest neighbor coupling is considered. A $p_z$-orbital coupling parameter of $t = 3\text{eV}$ was assumed.

Figure 3 shows that a zigzag nanotube is composed of rings of carbon atoms in the A- and B-atom sublattices.  Each ring in the A-atom sublattice is adjacent in the x-direction to a ring in the B-atom sublattice.  There are *n* carbon atoms in each ring and a total of *N* atoms in the entire channel.  The *N* x *N* Hamiltonian matrix for the whole nanotube channel is block tridiagonal,

$$\mathbf{H} = \begin{bmatrix} \alpha_1 & \beta_2^+ & & & & \\ \beta_2 & \alpha_2 & \beta_1 & & & \\ & \beta_1 & \alpha_3 & \beta_2 & & \\ & & \beta_2^+ & \alpha_4 & \beta_1 & \\ & & & \beta_1 & \alpha_5 & \cdots \\ & & & & \cdots & \cdots \end{bmatrix}, \tag{4}$$

where the *n* x *n* submatrix, $[\alpha_i]$, describes coupling within an A-type or B-type carbon ring, and the *n* x *n* $[\beta]$ matrices describe the coupling between adjacent rings.

In the nearest neighbor tight binding approximation, carbon atoms within a ring are uncoupled to each other so that $[\alpha_i]$ is a diagonal matrix. The value of a diagonal entry is the potential at that carbon atom site. If the nanotube is coaxially gated, the potential is invariant around the nanotube. The matrix, $[\alpha_i]$, therefore, is the potential at the *i*th carbon ring times the identity matrix, $[\alpha_i]=U_i[I]$.

There are two types of coupling matrices between nearest carbon rings, $[\beta_1]$ and $[\beta_2]$. As shown in Fig. 3, the first type, $[\beta_1]$, only couples an A(B) carbon atom to its B(A) counterpart in the neighboring ring. The coupling matrix is just the $p_z$ orbital coupling parameter times an identity matrix,

$$[\beta_1]=t[I]\ . \tag{5}$$

The second type of coupling matrix, $[\beta_2]$, couples an A(B) atom to two B(A) neighbors in the adjacent ring. The coupling matrix is

$$[\beta_2]=t\begin{bmatrix} 1 & & \dots & 1 \\ 1 & 1 & & \\ & 1 & 1 & \\ & & \dots & \dots \end{bmatrix}. \tag{6}$$

To understand eqn. (4), note that the odd numbered [α]'s refer to A-type rings and the even numbered one to B-type rings. Each A-type ring couples to the next B-type ring according to $[\beta_2]$ and to the previous B-type ring according to $[\beta_1]$. Each B-type ring couples to the next A-type ring according to $[\beta_1]$ and to the previous A-type ring according to $[\beta_2]$.

Having specified the Hamiltonian matrix for the channel, the next step is to compute the N x N self-energy matrices for the source and drain contacts, $[\Sigma_S]$ and $[\Sigma_D]$. The self-energies describe the open boundary conditions for the Schrödinger equation. Only the carbon atoms on the first and last rings couple to the contacts, so $[\Sigma_S]$ and $[\Sigma_D]$ are sparse, with a structure of

$$[\Sigma_S] = \begin{bmatrix} \Sigma_{11} & 0 & \dots & 0 \\ 0 & 0 & \dots & 0 \\ \vdots & \vdots & \ddots & \vdots \\ 0 & 0 & \dots & 0 \end{bmatrix}, \tag{7}$$

where $\Sigma_{11}$ is an *n* x *n* submatrix. Similarly, for $[\Sigma_D]$, the only nonzero block is the last diagonal submatrix. The derivation of these submatrices is described in detail in Appendix A.

The retarded Green's function,

$$\mathbf{G}^r = \left[(E + i0^+)\mathbf{I} - \mathbf{H} - \boldsymbol{\Sigma}_S - \boldsymbol{\Sigma}_D\right]^{-1}, \tag{8}$$

describes the bulk nanotube by **H** and the connection to the two contacts by the self-energy matrices. All matrices are size *N* x *N* with *N* being the total number of carbon atoms in the device. Solving eqn. (8) is equivalent to solving $\mathbf{A}\mathbf{G}^r = \mathbf{I}$, where $\mathbf{A} = \left[(E + i0^+)\mathbf{I} - \mathbf{H} - \boldsymbol{\Sigma}_S - \boldsymbol{\Sigma}_D\right]$. The straightforward but computationally approach is to compute $\mathbf{G}^r$ is by directly inverting the **A** matrix. Significant computational savings can be achieved by exploiting the block tridiagonal structure of **A**, which allows $\mathbf{G}^r$ to be computed by a recursive algorithm without inverting a large matrix [Svi02]. If the channel consists of $N_C$ carbon rings of a (*n*, 0) nanotube, the computational cost of directly inverting **A** goes as $\mathrm{O}[(n \times N_c)^3]$ whereas with the recursive algorithm it is only $\mathrm{O}(n^2 \times N_C)$. For the ballistic case, the solution is particularly efficient because only the first and last *n* columns of the Greens' function are needed.

Having computed the Green's function, the local density of states can be obtained, and the states can be filled according to the Fermi levels of the two contacts so that the charge density within the device can be found from eqn. (2). A method to compute the charge density from the Green's function using the recursive algorithm is also discussed in [Svi02]. By iterating between the NEGF equations to find the charge density and the Poisson equation to find the self-consistent potential, a self-consistent solution is obtained. The current is then evaluated from eqn. (3), where the current transmission probability, is obtained from the first diagonal block of the retarded Green's function,

$$\begin{aligned} T(E) &= \text{Trace}\left(\Gamma_S \mathbf{G}^r \Gamma_D \mathbf{G}^{r^+}\right) \\ &= \text{Trace}\left(\Gamma_{S(1,1)}\left\{i[G^r{}_{(1,1)} - G^r{}_{(1,1)}{}^+] - G^r{}_{(1,1)}\Gamma_{S(1,1)}G^r{}_{(1,1)}{}^+\right\}\right) \end{aligned} \tag{9}$$

where $\Gamma_{S,D} = i\left(\Sigma_{S,D} - \Sigma_{S,D}^+\right)/2$ is the source(drain) broadening and (1,1) denotes the first diagonal block of a matrix.

### 3.2 Mode space approach

The atomistic real space approach produces a matrix whose size is the total number of carbon atoms in the nanotube, which means that it is computationally intensive. A mode space approach significantly reduces the size of the Hamiltonian matrix. (A similar approach has been used for nanoscale MOSFETs [Ven02]). In brief, the idea is to exploit the fact that in a carbon nanotube, periodic boundary conditions must be applied around the circumference of the nanotube, so $k_C C = 2\pi q$, where $C$ is the circumference of the nanotube and $q$ is an integer. Transport may be described in terms of these circumferential modes. If $M$ modes contribute to transports, and if $M < n$, then the size of the problem is reduced from ($n$ x $N_C$) unknowns to ($M$ x $N_C$). If, in addition, the shape of the modes does not vary along the nanotube, then the $M$ circumferential modes are uncoupled, and we can solve $M$ one-dimensional problems of size, $N_C$, which is the number of carbon rings along the nanotube. Mathematically, we perform a basis transformation on the ($n$, 0) zigzag nanotube to decouple the problem into $n$ one-dimensional mode space lattices. The matrix is also tridiagonal, which allows the application of the efficient recursive algorithm for computing the Green's function [Svi02].

When a zigzag nanotube is coaxially gated, the modes around the tube are simple plane waves with wave vectors satisfying the periodic boundary condition, and the mode space approach exactly reproduces the results of the real space approach. The mathematical details for obtaining the Hamiltonian matrix for a mode are provided in Appendix B. A pictorial view is shown in Fig. 4. After the basis transformation, the two dimensional nanotube lattice is transformed to *n*, uncoupled one-dimensional lattices in mode space. As shown in Appendix B, the Hamiltonian matrix for the *q*th mode is

$$\mathbf{H}_q = \begin{bmatrix} U_1 & b_{2q} & & \\ b_{2q} & U_2 & t & \\ & t & U_3 & b_{2q} \\ & & & \cdots \end{bmatrix}, \tag{10}$$

where $U_i$ is the electrostatic potential at the *i*th carbon ring, $t$ is the C-C nearest neighbor binding parameter, and $b_{2q} = 2t\cos(\pi q/n)$. Equation (10) should be compared with eqn. (4). In eqn. (10), each element is a number, not an *n* x *n* submatrix as in eqn, (4). As in eqn. (4), the odd-numbered diagonal entries refer to the A-type submatrix and even numbered ones to the B-type submatrices. Each A-type ring couples to the next B-type ring with the parameter, $b_{2q}$ (analogous to $\beta_2$ in eqn. (4)) and to the previous B-type ring with the parameter, *t* (analogous to $\beta_1$ in eqn. (4)). Similarly, each B-type ring couple to the next A-type ring with parameter, *t*, and to the previous B-type ring with parameter, $b_{2q}$.

For an (*n*, 0) nanotube, there are *M* = *n* circumferential modes, but the computational cost is reduced when the modes are uncoupled. The computational cost can be further reduced by noticing that typically only one or a few modes are relevant to carrier transport. Modes with their band edges well above or below the source and drain Fermi levels are unimportant to carrier transports. The *E*(*k*) relation for the *q*th mode as computed from eqn. (10) is $E(k) = \pm\sqrt{t^2 + b_{2q}^2 + 2tb_{2q}\cos(3ka_{CC}/2)}$, where $a_{cc} \approx 1.42\,\text{Å}$ is the C-C bonding distance. The *q*th mode produces a conduction band and a valence band with symmetric *E*(*k*) , and a band gap of $E_g = 2|t|\{1+\cos(\pi q/n)\}$. When $n \bmod 3 = 0$, the lowest subband index is $q = 2n/3$, which results in $b_{2q} = -t$ and a zero band gap. Otherwise, the nanotube is semiconducting and the lowest subband index is the integer closest to $2n/3$. By retaining only those modes whose carrier population changes with device bias or operating temperature, the size of the problem is significantly reduced.

The mode space source and drain self-energies can be computed using the same recursive relation for the surface Green's functions already discussed in Appendix A. The details are provided in Appendix B. The structure of the self energy matrices is the same as in eqn. (7) except that $\Sigma_{11}$ (and $\Sigma_{NCNC}$ for the drain self energy) are numbers rather than *n* x *n* submatrices. After obtaining the Hamiltonian matrix and contact self energies, the retarded Green's function is computed. Because the Hamiltonian matrix for a mode is tridiagonal and only a small part of the retarded Green's function is needed for the purpose of computing charge density and current at the ballistic limit, the recursive algorithm [Svi02] or Gaussian elimination, rather than explicit matrix inversion, is used to compute the retarded Green's function.

## 4. Phenomenological Treatment of Metal/CNT junctions

In carbon nanotube transistors, the metal source and drain are typically attached directly to the intrinsic nanotube channel, and the gate modulates the source-drain current by changing the transmission through the Schottky barrier at the source end of the channel. To properly simulate such devices, the metal/CNT junction must be treated quantum mechanically. We currently treat this problem phenomenologically by defining an appropriate self-energy. Note that the self-energies defined in Sec. 3 do not apply here – they assume that carriers enter and leave the device without the need to tunnel through any barriers at the contact. As shown in Fig. 5, the phenomenological self energy must contain two parameters, one to describe the barrier height and another the density of metal-induced gap states (MIGS). Our approach mimics the effect of a real metal contact by specifying its work function and by injecting a continuous density of states near the Fermi level. This approach has proven useful in understanding transistor operations of Schottky barrier CNTFETs [Guo03a].

The phenomenological treatment is described in Appendix C. In brief, each semiconducting mode in the semiconducting zigzag nanotube is coupled at the M/CNT interface to a mode of a metallic zigzag CNT. As shown in Appendix C, $\Sigma_{11}$ in eqn. (7) becomes

$$\Sigma_{MS} = \alpha t^2 g_S = \alpha \frac{E - E_{m1} - \sqrt{(E - E_{m1})^2 - 4t^2}}{2}. \tag{11}$$

The coupling is described by two parameters. The first parameter is $\phi_{bo}$, the Schottky barrier height for electrons without the presence of the interface states, which describes the band discontinuity at the interface and provides the value for $E_{m1}$, the mid-gap energy of the CNT at the interface. ($E_{m1} = E_{Fm} + \phi_{b0} - E_g / 2$, where $E_{Fm}$ is the metal Fermi level and $E_g$ is the CNT band gap.) The second parameter is the tight-binding parameter, $\alpha$, between the semiconducting and the metallic mode ($0 < \alpha \leq 1$), which determines how well the metal contact is coupled to the nanotube channel, and is roughly proportional to the density of metal-induced-gap-states (MIGS). This simple model describes the interface at a level similar to those in the literature that the band discontinuity and density of interface states as input parameters [Leo00].

## 5. The Overall Simulation Procedure

The overall simulation must be done self-consistently with Poisson's equation. Figure 6 shows the modeled, coaxial gate CNTFET, which provides the theoretically best gate control over the channel [Aut97]. The source and drain are heavily doped, semi-infinite carbon nanotubes, and the gate modulates the conductance of the channel, just like in a conventional Si MOSFET. For this device, we use the self energies described in Appendix A or Appendix B. By using a self-energy for metal/NT contacts as discussed in Appendix C, the simulation scheme can also be applied to Schottky barrier CNTFETs.

The transistor I-V characteristics strongly depend on the interplay of quantum transport and electrostatics, so we performed a self-consistent iteration between the NEGF transport equation and the Poisson equation as shown in Fig. 7. In brief, the procedure is as follows. For a given charge density, the Poisson equation is solved to obtain the electrostatic potential in the nanotube channel. Next, the computed potential profile is used as the input for the NEGF transport equation, and an improved estimate for the charge density is obtained. The iteration between the Poisson equation and the NEGF transport equation continues until self-consistency is achieved. Finally, the current for the self-consistent potential profile is computed.

For the coaxially gated carbon nanotube transistor, it is convenient to solve Poisson's equation in cylindrical coordinates. Since the potential and charge density are invariant around the nanotube, the Poisson equation is essentially a 2D problem along the tube (*x*-direction) and the radial direction (*r*-direction) as shown in Fig. 6. Poisson's equation is written as

$$\nabla^2 E_m(r,z) = -\frac{e}{\varepsilon}\rho \, , \tag{11}$$

where $E_m$ is defined as the vacuum energy level minus the work function of an intrinsic nanotube, and is exactly the middle gap energy for the grid points on the tube surface, and $\rho$ is the charge density, which is non-zero only for grid points on the tube surface. The boundary condition applied at $r = 0$ is that the electric field along the *r*-direction is zero [Aut97],

$$\varepsilon_r \,|_{r=0} = 0 \, . \tag{12}$$

The potential at the gate electrode is known, so using the Fermi level of a grounded electrode as the zero energy, the electron potential at the gate electrode is,

$$E_m(gate) = -eV_G + \phi_{ms} \, , \tag{13}$$

where $V_G$ is the gate bias, and $\phi_{ms}$ is the work function difference between the gate metal and the intrinsic nanotube channel. By simulating a sufficiently large area, as shown in Fig. 6, Neumann boundary condition, which assumes that the electric field in the direction normal to the boundary is zero, can be applied to the remaining boundaries.

The continuous form of the Poisson equation, eqn. (11), is discretized for computer simulation. It is convenient to take a volume element near a grid point, as shown in Fig. 6, and apply the integral form of the Poisson equation to that volume element, which is a ring around the tube axis with a rectangular cross section,

$$\oint \vec{D} \cdot d\vec{S} = q_{ij} \ , \tag{13}$$

where $q_{ij}$ is the charge in the total volume element, which is non-zero only on tube surface. The discretized equation for an element at the grid point ($x_i$, $r_j$) in air, is

$$2\pi\varepsilon_0 \left( \frac{r_{j-1} + r_j}{2} \Delta x \frac{E_m^{i,j-1} - E_m^{i,j}}{\Delta r} + \frac{r_{j+1} + r_j}{2} \Delta x \frac{E_m^{i,j+1} - E_m^{i,j}}{\Delta r} + r_j \Delta x \frac{E_m^{i+1,j} - E_m^{i,j}}{\Delta r} + r_j \Delta x \frac{E_m^{i-1,j} - E_m^{i,j}}{\Delta r} \right) = e\Delta x (N_D - n_{net}) \ . \tag{14}$$

For grid points in the gate insulator, the gate insulator dielectric constant replaces $\varepsilon_0$ in eqn. (14). For the grid points at the gate insulator/air interface, the air dielectric constant is used for volume surfaces in air and the gate insulator dielectric constant is used for volume surfaces in the gate insulator.

Equation (14) is linear and mathematically easy to solve, but the convergence of the quantum transport and the linear Poisson equation is poor [Ren01]. A non-linear Poisson equation, which relates the charge density to the potential through a non-linear dummy function, has been proven to be very useful in improving the convergence. The non-linear dummy function relating the charge density and the potential should be as close to the physical relation determined by carrier transport equation as possible for better convergence. Typically, semiclassical, equilibrium carrier statistics with a dummy quasi Fermi level are used as the dummy function. The non-linear Poisson equation takes the charge density computed by the transport equation as the input, and converts the charge density to a quasi Fermi level using the dummy function. Then the non-linear Poisson equation is solved for the potential by Newton-Ralphson iteration. Details of the non-linear Poisson solver can be found in [Ren01].

## 6. Results

The simulation methods discussed in the previous sections have proven useful in several recent transistor studies [Guo03a, b]. The purpose of this section is to show some simple examples to demonstrate that: i) quantum effects are captured, ii) the mode space approach is valid when potential is uniform around the tube, and iii) the metal/CNT junction can be treated by our phenomenological self energy.

We first simulate a coaxially gated, MOSFET-like CNTFET as shown in Fig. 6. The transistor channel is a (25,0) intrinsic CNT, which results in a band gap of ~0.42eV and a diameter of ~2nm. The nanotube length is ~50nm, consisting of ~$1.2\times10^4$ carbon atoms. A self-consistent Poisson-NEGF simulation in the real space (using the recursive algorithm for computer the Green's function) is performed. Fig. 8a shows the energy-resolved local-density-of-states (LDOS), and the energy band profile. The band gap region with extremely low LDOS (darker in the grayscale plot) can be clearly identified. Due to the existence of the barriers, the source/drain incident wave is reflected and the quantum interference pattern between the incident and reflected waves is apparent. A quantum well is formed in the valence band of the channel, and the 1$^{st}$ and 2$^{nd}$ confined states with one or two LDOS maxima, respectively, can be clearly seen. The band edge of the second subband is also observed. Figure 8b shows the energy resolved electron density (electron density spectrum), which is obtained by filling the LDOS with the source or drain Fermi level. The bandgap, quantum interference, quantum confinement, and the second subband can still be clearly seen.

Next, we explore the validity of the mode space approach by comparing the results of the real space approach to those of the mode space approach. The mode space approach theoretically should exactly reproduce the results of the real space approach when the potential is invariant around the tube, and a sufficient number of modes is included in the mode space simulation. A CNTFET as shown in Fig. 6 with a (13,0) nanotube channel, which results in a band gap of ~0.83eV and a diameter of ~1nm, is simulated. The carbon nanotube length is ~50nm, consisting of ~6000 carbon atoms. Because the third subband is ~1eV away from the lowest subband and the applied bias is $\le 0.4$ V, only the lowest two subbands are treated in the mode space simulation. The Hamiltonian matrix for the lowest subband is small ($\sim 500\times500$), and computing the Green's function for a subband using the recursive algorithm is fast even on a single CPU PC. Figure 9, which compares the I-V characteristics of the real space and mode space approaches, shows that the mode space simulation excellently reproduces the results of the real space approach. Figure 10, which plots the band profile and the charge density at on-state, again shows that the mode space approach excellently reproduces the results from the real space approach results. The good agreement between the real and mode space approach results from the equal potential around the tube direction when it's coaxially gated. The mode space is highly advantageous in reducing the computational burden, and it is valid when the potential variation around the tube is much smaller than the spacing between the subbands.

Finally, we treat an SBFET-like CNTFET by self-consistent, quantum simulation. Fig. 11a shows the simulated transistor structure. The metal source/drain is directly attached to a (13,0) intrinsic nanotube channel, so a Schottky barriers forms between the source/drain and channel. A mid-gap Schottky barrier, with equal barrier height for electrons and holes, is simulated. Fig. 11b shows the local density of states at $V_D = V_G = 0.4$ V. The metal-induced gap states (MIGS) near the metal/CNT interfaces are apparent and decay rapidly with a tail of a few nanometers inside the channel. The tunneling states under the Schottky barrier in the conduction band at the source end of the channel are clear. The metal-nanotube interface is not perfectly transmitting, and the weakly confined states with the increasing number of LDOS maxima, due to the weak localization created by double metal/CNT barriers at the source and drain ends of the channel, can be seen. The atomistic-scale oscillations of the charge density spectrum along the channel direction is probably due to the charge transfer between A and B types of carbon rings in a zigzag carbon nanotube [Leo02].

## 7. Discussion

The semi-empirical approach described in this paper is only one part of a multi-scale hierarchy shown in Fig. 1. More rigorous, *ab initio* methods are needed to treat the metal/CNT interface properly. Such simulations would allow first principles calculations of the barrier height and the MIGS, two parameters that we now treat as phenomenological. Such simulations may also provide useful insights into how to produce ohmic, rather than Schottky barrier, contacts when desired. The phenomenological model, however, is well-suited for device-scale simulations because of its computational efficiency. One approach to this multi-scale challenge is to relate the phenomenological parameters for the metal/nanotube contacts in the semi-empirical approach described in this paper to detailed atomistic simulations of the contact. In such an approach, the semi-empirical model would stand alone and be related to separate, *ab initio* simulations. Another possibility is the domain decomposition approach sketch in Fig. 1. In this approach, the key challenge is to connect the two regions, described with much different sets of basis functions, through the self-energies. This "mixed basis set approach" is already being applied to problems involving molecules on silicon contacts [Rak03] and is being investigate for the metal/nanotube contact as well.

The approach described in this paper assumes ballistic transport, but scattering by phonon emission is likely to be a factor in devices under realistic operating voltages [Yao00, Jav03b, Par03]. There is a clear prescription for treating the electron-phonon interaction within the NEGF formalism [Dat95], but the computational burden increases rapidly. It is likely, therefore, that semiclassical, continuum approaches like those used to treat semiconductor devices by Monte Carlo simulation [Fis88] will be needed when a detailed treatment of the electron-phonon coupling is necessary. NEGF simulation is the method of choice when quantum transport is the dominant factor, and phenomenological treatments of scattering can be used [Dat00]. Semiclassical approaches are the method of choice when scattering dominates, and phenomenological quantum corrections can be made.

Finally, work at the device level needs to be coupled to circuit level models so that the system level implications of novel devices can be readily explored. Existing approaches may or may not be adequate. CNTFETs, for example, should operate near the ballistic limit, and it is not clear that traditional MOSFET models, which were developed for the scattering-dominated regime, can be extended to quasi-ballistic transistors. Recently, a new circuit model for ballistic CNTFETs has been developed [Ari03]. The more general question of how circuit models for new exploratory devices can be rapidly developed is an important one to address.

## 8. Conclusions

Methods for the NEGF/Poisson simulation of carbon nanotube transistors were discussed and illustrated. The real space approach, which uses one $p_z$ orbital per carbon atom as the basis, achieves atomistic resolution for quantities of interest. Significant computational saving can be achieved by using the mode space approach, which performs a basis transformation around the nanotube circumferential direction and transforms the 2D nanotube lattice to decoupled 1D mode space lattices. Each mode in the mode space approach describes one conduction subband and its corresponding valence subband, and atomistic resolution along the transport direction is retained. The simulation methods discussed in this paper have been applied to several transistor studies with the purpose of understanding experiments and exploring device physics [Guo03a, 03b]. Finally, the need to complement this semi-empirical device level model with higher level circuit models and lower level *ab initio* models was discussed.

### Acknowledgement

The authors are indebted to Dr. Mani Vaidyanathan, Dr. Diego Kienle, Dr. Avik Ghosh, and Sayed Hasan of Purdue University for helpful discussions. This work was supported by the NSF Network for Computational Nanotechnology, grant no. EEC-0228390, and the MARCO Focused Research Center on Materials, Structure, and Devices.

## Appendix A. The source/drain self-energies the real space

The overall size of the self-energy matrices for the source and drain contacts is the same as the Hamiltonian matrix for the channel, but the self-energy matrices are highly sparse. For example, only one carbon ring at the source end of the channel couples to the source, thus only one submatrix, the (1,1) submatrix in the basis used for eqn. (4), is non-zero for the source self energy, $\Sigma_1$. Similarly, only one submatrix is non-zero for the drain self-energy, $\Sigma_2$. The non-zero entry of the self-energies can be computed by a recursive relation for the surface Green's function, with details explained in the appendix of [Ven02]. Here we compute the self-energy for a semi-infinite nanotube source. The self-energy approach can be readily extended to treat any type of contacts, for example, metal-nanotube contacts, as will be discussed later.

Fig. A1 shows how carbon rings are coupled for a semi-infinite nanotube source. Each circle (triangle) represents a carbon ring consisting of A(B)-type carbon atoms. The carbon ring couples to the nearest ring, with a coupling matrix of $\beta_1$ or $\beta_2$, and $g_m$ is the surface Green's function for the *m*th ring in the source extension, ordered from the source/channel interface. The recursive relation [Ven02] relates the surface Green's functions,

$$g_m = [(E + i0^+)I - \alpha_m - \tau g_{m+1}\tau^+]^{-1}, \tag{A1}$$

where $\tau$ is the coupling matrix between the *m*th and the *(m+1)*th carbon rings and $\alpha_m$ is the Hamiltonian matrix of the *m*th ring. Applying this recursive relation to the nanotube in Fig. A1, we get

$$\begin{aligned} g_1 &= [(E + i0^+)I - \alpha_1 - \beta_2^+ g_2 \beta_2]^{-1} \\ g_2 &= [(E + i0^+)I - \alpha_2 - \beta_1^+ g_3 \beta_1]^{-1} \end{aligned} \tag{A2}$$

Note that the potential is invariant inside the source, so $\alpha_1 = \alpha_2$. Furthermore, $g_1 = g_3$ due to the periodicity of the nanotube lattice. Using these relations, eqn. (A2) becomes two coupled matrix equations with two unknowns, $g_1$ and $g_2$. The surface Green's function can be numerically solved from Eqn (A2). The non-zero submatrix of the source self-energy matrix is $\Sigma_S^{1,1} = \beta_1 g_1 \beta_1^+$, where the superscript denotes that it is the (1,1) submatrix of the overall source self-energy matrix. The self-energy for the drain contact can be computed in a similar way.

## Appendix B. The transistor Hamiltonian in mode space

The following basis transformation, which transforms the real space basis around the nanotube to a mode space basis, is performed to the Hamiltonian matrix in the real space,

$$H^{'} = \begin{bmatrix} V & & & & & \\ & V & & & & \\ & & V & & & \\ & & & V & & \\ & & & & V & \\ & & & & & ... \end{bmatrix}^{+} \begin{bmatrix} \alpha_1 & \beta_2^{+} & & & & \\ \beta_2 & \alpha_2 & \beta_1 & & & \\ & \beta_1 & \alpha_3 & \beta_2 & & \\ & & \beta_2^{+} & \alpha_4 & \beta_1 & \\ & & & \beta_1 & \alpha_5 & ... \\ & & & & ... & ... \end{bmatrix} \begin{bmatrix} V & & & & & \\ & V & & & & \\ & & V & & & \\ & & & V & & \\ & & & & V & \\ & & & & & ... \end{bmatrix}$$

$$= \begin{bmatrix} \alpha_1' & \beta_2^{'+} & & & & \\ \beta_2' & \alpha_2' & \beta_1' & & & \\ & \beta_1' & \alpha_3' & \beta_2' & & \\ & & \beta_2^{'+} & \alpha_4' & \beta_1' & \\ & & & \beta_1' & \alpha_5' & ... \\ & & & & ... & ... \end{bmatrix} \tag{B1}$$

with

$\alpha_i^{'} = V^{+}\alpha_i V$ ,

$\beta_1^{'} = V^{+}\beta_1 V$ , and

$$\beta_2^{'} = V^{+}\beta_2 V , \tag{B2}$$

where V is the transform matrix from the real space basis of a carbon atom ring to the mode space basis. Our purpose is to uncouple the modes after the basis transformation, *i.e.*, to make the Hamiltonian matrix elements between different modes equal to zero. This requires that after the transformation, $\alpha_i^{'}$, $\beta_1^{'}$, and $\beta_2^{'}$, become diagonal matrices.

Notice that $\alpha_i$ or $\beta_1$ is a constant times an identify matrix. These matrices remain unchanged and diagonal after any basis transformation,

$$\begin{aligned} \alpha_i^{'} &= \alpha_i = U_i I \\ \beta_1^{'} &= \beta_1 = tI \end{aligned} . \tag{B3}$$

The problem now becomes to find out the eigenvectors and corresponding eigenvalues for

$$\beta_2 = t\begin{bmatrix} 1 & & \dots & 1 \\ 1 & 1 & & \\ & 1 & 1 & \\ & & \dots & \dots \end{bmatrix}. \tag{B4}$$

The $q$th eigenvectors of $\beta_2$ is the plane wave around the nanotube

$$\psi_q(n_y) = \frac{1}{\sqrt{n}} e^{ik_q n_y} , \tag{B5}$$

where the qth wave vector, $k_q$, satisfies the periodic boundary condition, $k_q = 2\pi q / n \, [\, 0 \le q \le (n-1)\, ]$, and the qth eigenvalue is $b_{2q} = 2te^{-\pi q i/n} \cos(\pi q/n)$. After the basis transformation, $\beta_2$ becomes

$$\beta_2^{'} = V^{+} \beta_2 V = \begin{bmatrix} b_{21} & & & \\ & \dots & & \\ & & b_{2q} & \\ & & & \dots \end{bmatrix}. \tag{B6}$$

All submatrices, $\alpha', \beta_1^{'}$, and $\beta_2^{'}$ are diagonal, and there are no matrix elements between different modes around the nanotube after the basis transformation. If we reorder the basis according to the modes, the Hamiltonian matrix is

$$
H^{'} = \begin{bmatrix} H_1^{'} & & & & \\ & H_2^{'} & & & \\ & & \cdots & & \\ & & & H_q^{'} & \\ & & & & \cdots \end{bmatrix}, \tag{B7}
$$

where $H_q^{'}$ is the Hamiltonian matrix for the *q*th mode,

$$
H_q^{'} = \begin{bmatrix} U_1 & b_{2q}^{+} & & & \\ b_{2q} & U_2 & t & & \\ & t & U_3 & b_{2q} & \\ & & & \cdots & \end{bmatrix}, \tag{B8}
$$

and all off-diagonal submatrices of *H'* are zero because all modes are decoupled after the basis transformation. Each mode can be separately treated in the mode space, thus the Hamiltonian matrix size is greatly reduced. Furthermore, for typical terminal bias conditions, only a few modes are relevant to carrier transport, which further reduces the computational load. The phase factor of the complex number, $b_{2q}$, has no effect on the results such as charge density and current, thus it can be omitted and $b_{2q} = 2t\cos(\pi q/n)$ can be used instead.

The source and drain self-energies in the mode space can be computed using the same recursive relation for the surface Green's function as already shown in Appendix A. For the qth mode, the matrix $\beta_1$ in eqn. (A2) is replaced by $t$, and $\beta_2$ is replace by $b_{2q}$,

$$
\begin{aligned}
g_{1q} &= [(E + i0^{+})I - U_1 - b_{2q}^{+} g_{2q} b_2]^{-1}, \\
g_{2q} &= [(E + i0^{+})I - U_1 - b_1^{+} g_{1q} b_1]^{-1},
\end{aligned} \tag{B9}
$$

where $g_{1q}$ and $g_{2q}$ are the surface Green's functions for the first and second node inside the source as shown in Fig. A1, and $U_1$ is the source potential. The recursive equations in the mode space are number equations and can be analytically solved,

$$g_{1q} = \frac{(E-U_1)^2 + b_1^2 - b_{2q}^2 \pm \sqrt{\left[(E-U_1)^2 + b_1^2 - b_{2q}^2\right]^2 - 4(E-U_1)^2 b_1^2}}{2b_1^2(E-U_1)}. \qquad \text{(B10)}$$

The retarded surface Green's function for the first node inside the source, $g_{1q}^r$, is the one with the negative imaginary part, and the source self-energy for the *q*th mode is $\Sigma_{Sq} = b_1^2 g_{1q}$.

## Appendix C. Phenomenological treatment of metal-nanotube contacts

The metal/CNT junction is treated in the atomistic mode space. The *q*th mode of a semiconducting, zigzag CNT is

$$H_{CNT} = \begin{bmatrix} E_{m1} & b_{2q} & & & \\ b_{2q} & E_{m2} & t & & \\ & t & E_{m3} & b_{2q} & \\ & & b_{2q} & E_{m4} & \ddots \\ & & & \ddots & \ddots \end{bmatrix}, \tag{D1}$$

where $E_{mi}$ is the middle gap potential at the *i*th carbon ring. To mimic the continuous states injected from the metal to the semiconducting nanotube, each semiconducting mode is coupled at the M/CNT interface to the metallic mode of metallic zigzag CNTs, which has a constant density of states over a large energy range. The Hamiltonian for the metallic subband is

$$H_{met} = \begin{bmatrix} \ddots & \ddots & & \\ \ddots & E_{m1} & t & \\ & t & E_{m1} & -t \\ & & -t & E_{m1} \end{bmatrix}, \tag{D2}$$

where $E_{m1}$ is the energy of the crossing point of the metallic bands, and is the same as $E_{m1}$ in eqn. (D1) if the mid-gap energy is assumed to be continuous at the interface for simplicity. (The simulation results are insensitive to the value of $E_{m1}$ in eqn. (D2) due to the nearly constant density-of-states near the Fermi point of the metallic bands.) The overall Hamiltonian matrix for the metal/CNT junction is

$$H = \begin{bmatrix} H_{met} & t\sqrt{\alpha} \\ t\sqrt{\alpha} & H_{CNT} \end{bmatrix} = \left[\begin{array}{cccc|cccc} \ddots & \ddots & & & & & & \\ \ddots & E_{m1} & t & & & & & \\ & t & E_{m1} & -t & & & & \\ & & -t & E_{m1} & t\sqrt{\alpha} & & & \\ \hline & & & t\sqrt{\alpha} & E_{m1} & b_{2q} & & \\ & & & & b_{2q} & E_{m2} & t & \\ & & & & & t & E_{m3} & \ddots \\ & & & & & & \ddots & \ddots \end{array}\right]. \tag{D3}$$

The metal contact is treated by computing its self-energy to the semiconducting channel. We again use the recursive relation for the surface Green's function of the metal contact,

$$g_S = [(E + i0^+) - E_{m1} - t g_S t]^{-1} \tag{D4}$$

with the solution,

$$g_S = \frac{E - E_{m1} - \sqrt{(E - E_{m1})^2 - 4t^2}}{2t^2}. \tag{D5}$$

The solution with a negative imaginary part is the retarded surface Green's function. The self-energy for the metal /CNT contact computed from the surface Green's function is

$$\Sigma_{MS} = \alpha t^2 g_S = \alpha \frac{E - E_{m1} - \sqrt{(E - E_{m1})^2 - 4t^2}}{2}. \tag{D6}$$

# FIGURES

Fig. 1 An illustration of how continuum, *ab initio*, atomistic and semi-empirical atomistic models will be combined in a multi-scale description of a carbon nanotube electronic device.

Fig. 2 The generic transistor with a molecule or device channel connected to the source and drain contacts. The source-drain current is modulated by a third electrode, the gate. The quantities in the NEGF calculation are also shown.

Fig. 3 The schematic diagram of a (*n*, 0) zigzag nanotube (*n* = 6 in this case). The circles are the A-type carbon atom sublattice, and the triangles are the B-type carbon atom sublattice. The coordinate system is also shown: c is the circumferential direction, and x is the carrier transport direction.

Fig. 4 (a) The real space 2D lattice of the (n,0) zigzag nanotube (b) The uncoupled, 1D lattices mode space lattices. A basis transformation from the real space to the *k* space is performed around the tube from (a) to (b).

Fig. 5 (a) The metal-carbon nanotube junction. (b) The band diagram of the junction. $E_C$, $E_V$ and $E_m$ are the conduction band edge, the valence band edge, and the middle gap energy in the nanotube, respectively. $E_{Fm}$ is the metal Fermi level, and $\phi_{bn}$ is the Schottky barrier height for electrons.

Fig. 6 The modeled, coaxially gated carbon nanotube transistor with heavily-doped, semi-infinite nanotubes as the source/drain contacts. The channel is intrinsic and the gate length equals the channel length. Also shown are the simulated area, the simulation grid and the cylindrical coordinate system used for solving the Poisson equation. The dashed rectangular area shows the volume element used to discretize the Poisson equation at ($x_i$, $r_j$).

Fig. 7 The self-consistent iteration between the NEGF transport and the electrostatic Poisson equation. These two equations are iteratively solved until self-consistency is achieved. Then the current is computed using the self-consistent potential.

Fig. 8 (a) The local-density-of-states (LDOS) and (b) the electron density spectrum computed by the real space approach at $V_G$=0.25V and $V_D$=0.4V. The modeled transistor is shown in Fig. 7a. The nanotube is a (25,0) CNT with a diameter d~2nm and bandgap Eg~0.4eV.

Fig. 9 The I-V characteristics computed by the real space approach (the solid line) and the mode space approach with 2 subbands (the circles) for a CNTFET as shown in Fig. 7. The (13,0) nanotube channel length is 15nm.

Fig. 10 (a) The conduction band profile computed by the real space approach (the solid lines) and the mode space approach (the circles) at $V_G = V_D = 0.4\,\text{V}$. (b) The charge density computed by the real space approach (the solid line) and the mode space approach (the dashed line). The solid and dashed lines lie on top of each other.

Fig.11 (a) The coaxially gated Schottky barrier carbon nanotube transistor with an intrinsic nanotube channel directly attached to metal source and drain contacts. The nanotube channel is a (13,0) zigzag CNT with a diameter d~1nm and band gap Eg~0.83eV. The gate insulator is a 2nm-thick $ZrO_2$. (b) The local-density-of-states (LDOS) at $V_D = V_G = 0.4\,\text{V}$, which clearly shows tunneling through the Schottky barrier at the source end of the channel, and metal induced gap states (MIGS) at the metal/CNT interfaces.

Fig.A1 Computing the source self-energy for a zigzag nanotube. The circles represent A-type carbon rings and the triangles represent B-type carbon rings. $g_i$ is the surface Green's function for the *i*th carbon ring inside the source. $\beta_1\,(\beta_2)$ is the first (second) kind coupling matrix between neighboring rings, as described in the text.

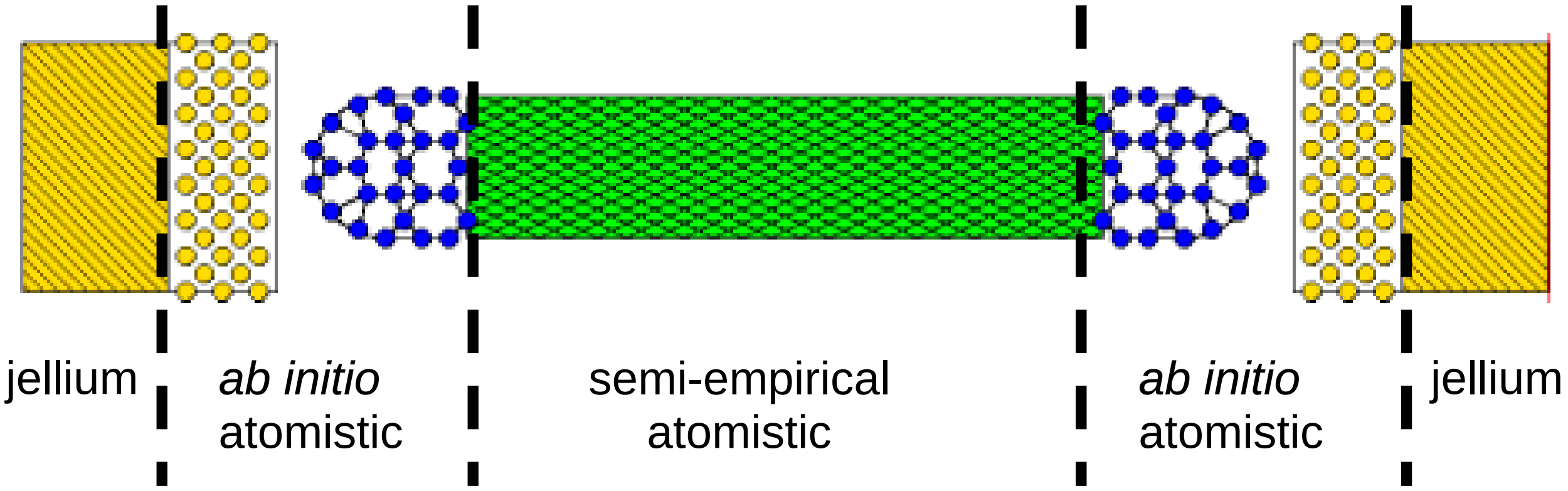

Fig. 1 An illustration of how continuum, *ab initio*, atomistic and semi-empirical atomistic models will be combined in a multi-scale description of a carbon nanotube electronic device.

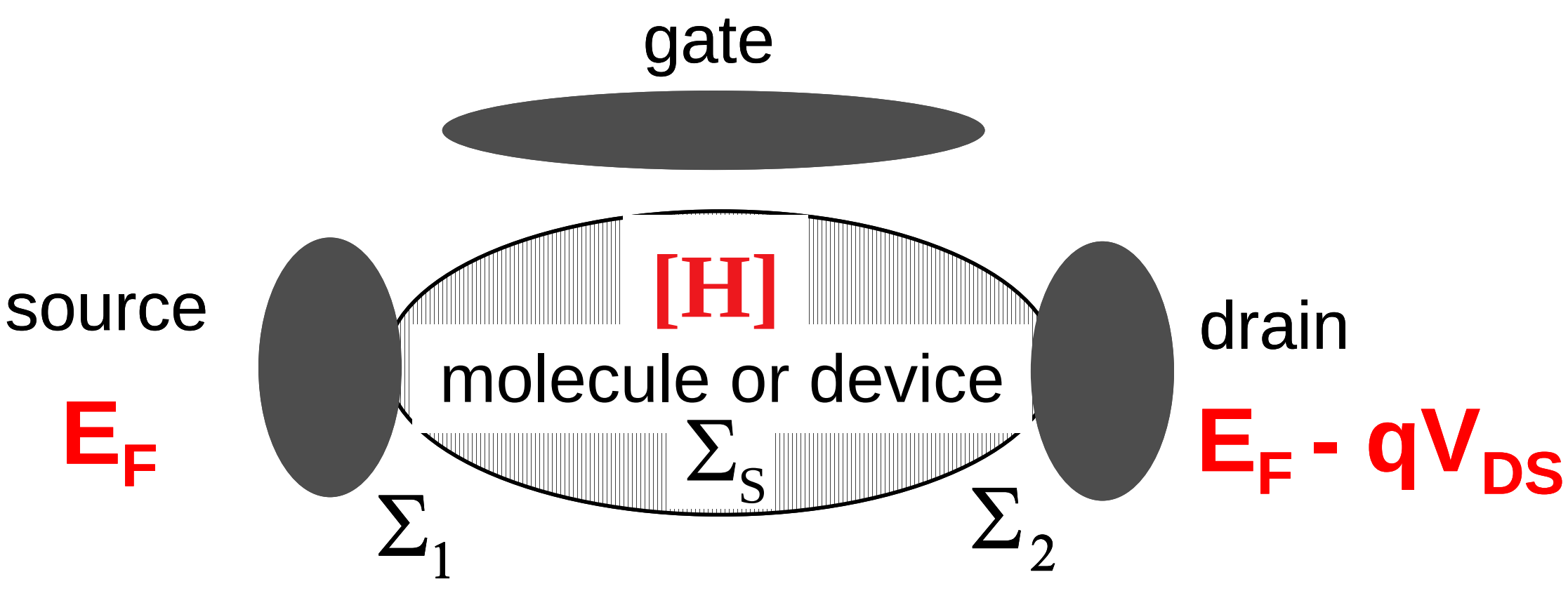

Fig. 2 The generic transistor with a molecule or device channel connected to the source and drain contacts. The source-drain current is modulated by a third electrode, the gate. The quantities in the NEGF calculation are also shown.

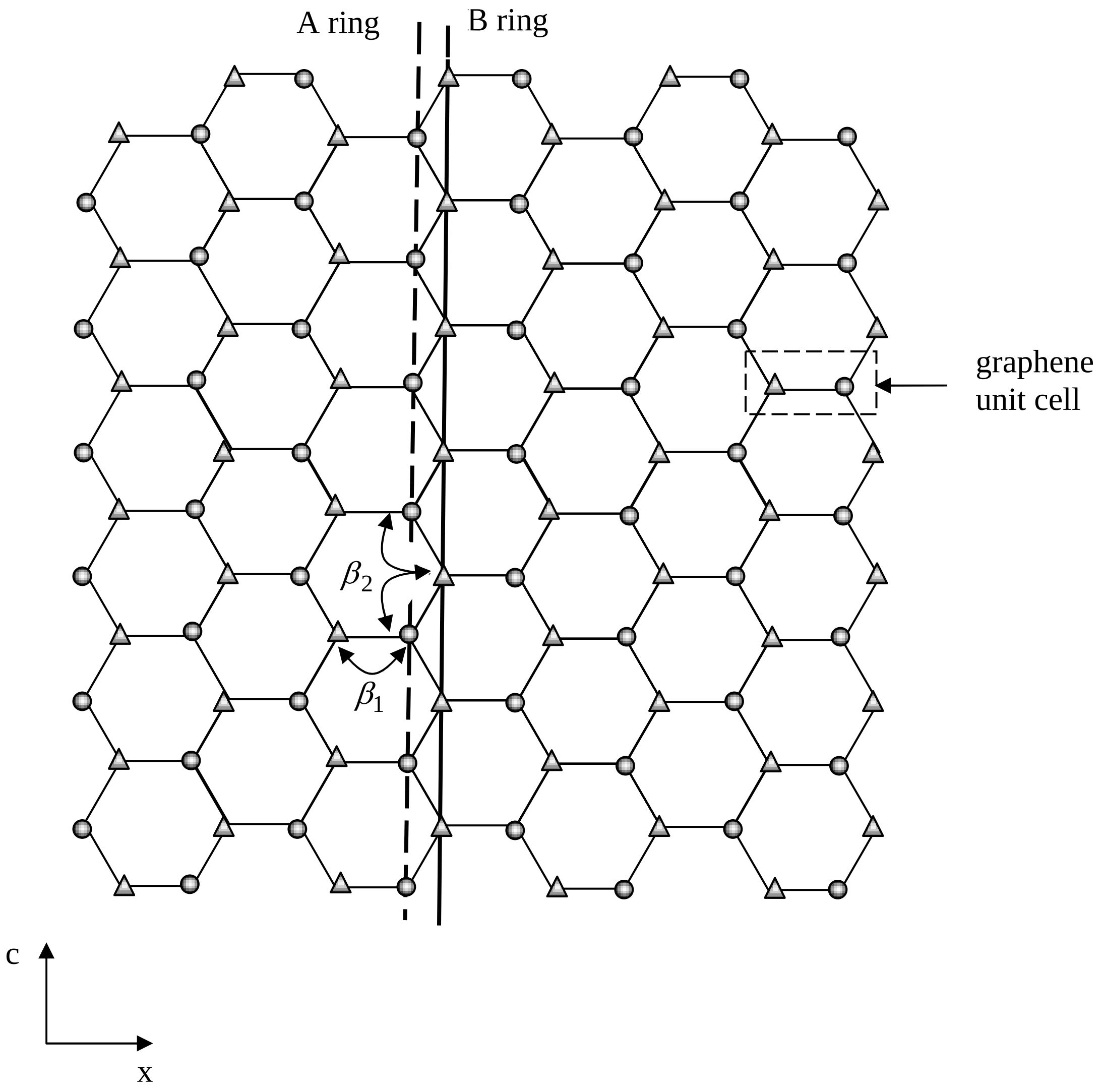

Fig. 3 The schematic diagram of a (*n*, 0) zigzag nanotube (*n* = 6 in this case). The circles are the A-type carbon atom sublattice, and the triangles are the B-type carbon atom sublattice. The coordinate system is also shown: c is the circumferential direction, and x is the carrier transport direction.

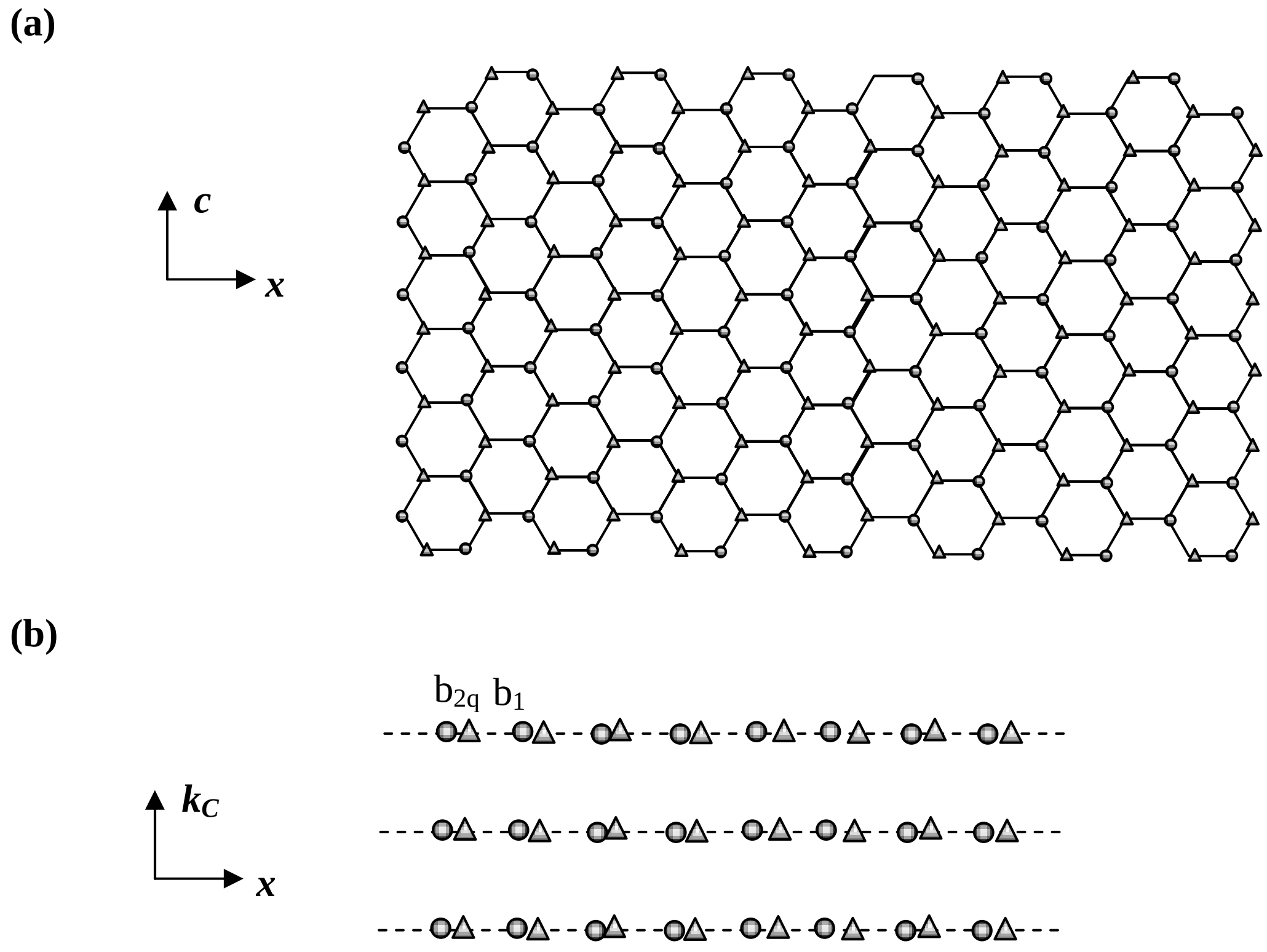

Fig. 4 (a) The real space 2D lattice of the (n, 0) zigzag nanotube (b) The uncoupled, 1D mode space lattices. A basis transformation on the real space lattice of (a) transforms the problem to the *M* one-dimensional problems, where M labels a specific $k_C$.

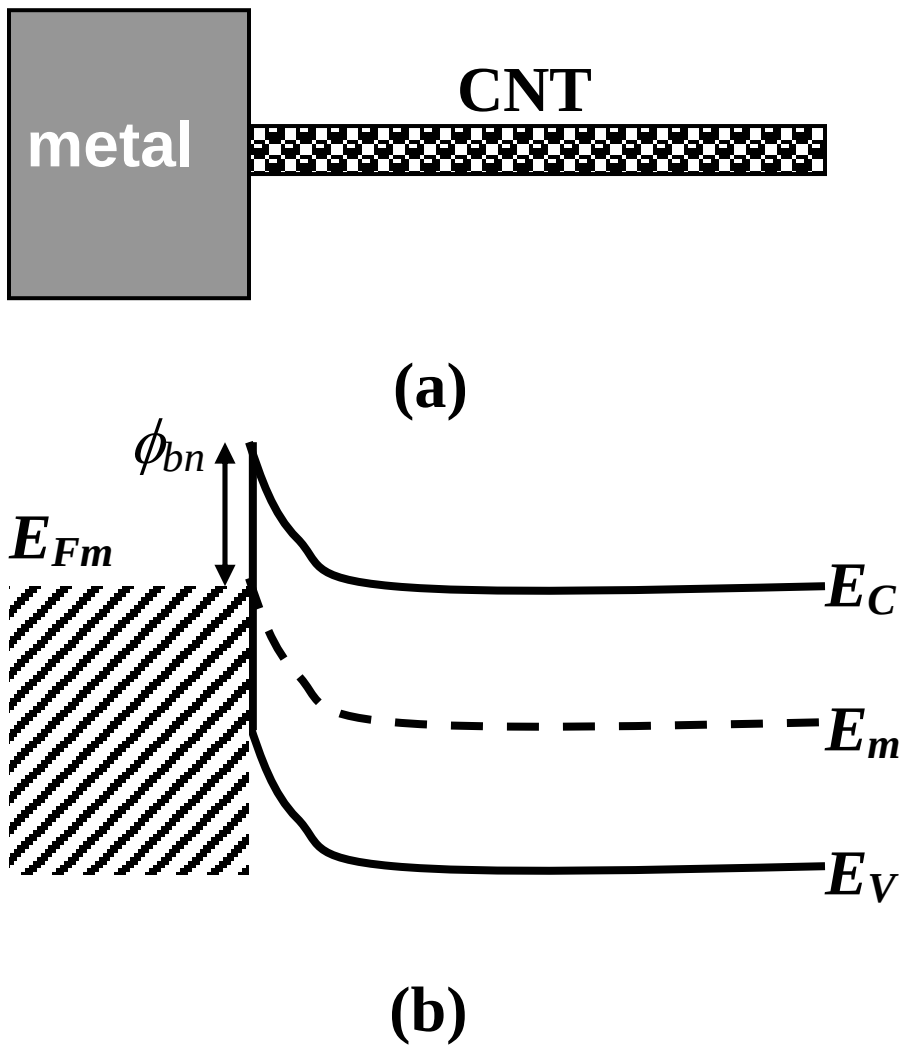

Fig. 5 (a) The metal-carbon nanotube junction. (b) The band diagram of the junction. $E_C$, $E_V$ and $E_m$ are the conduction band edge, the valence band edge, and the middle gap energy in the nanotube, respectively. $E_{Fm}$ is the metal Fermi level, and $\phi_{bn}$ is the Schottky barrier height for electrons.

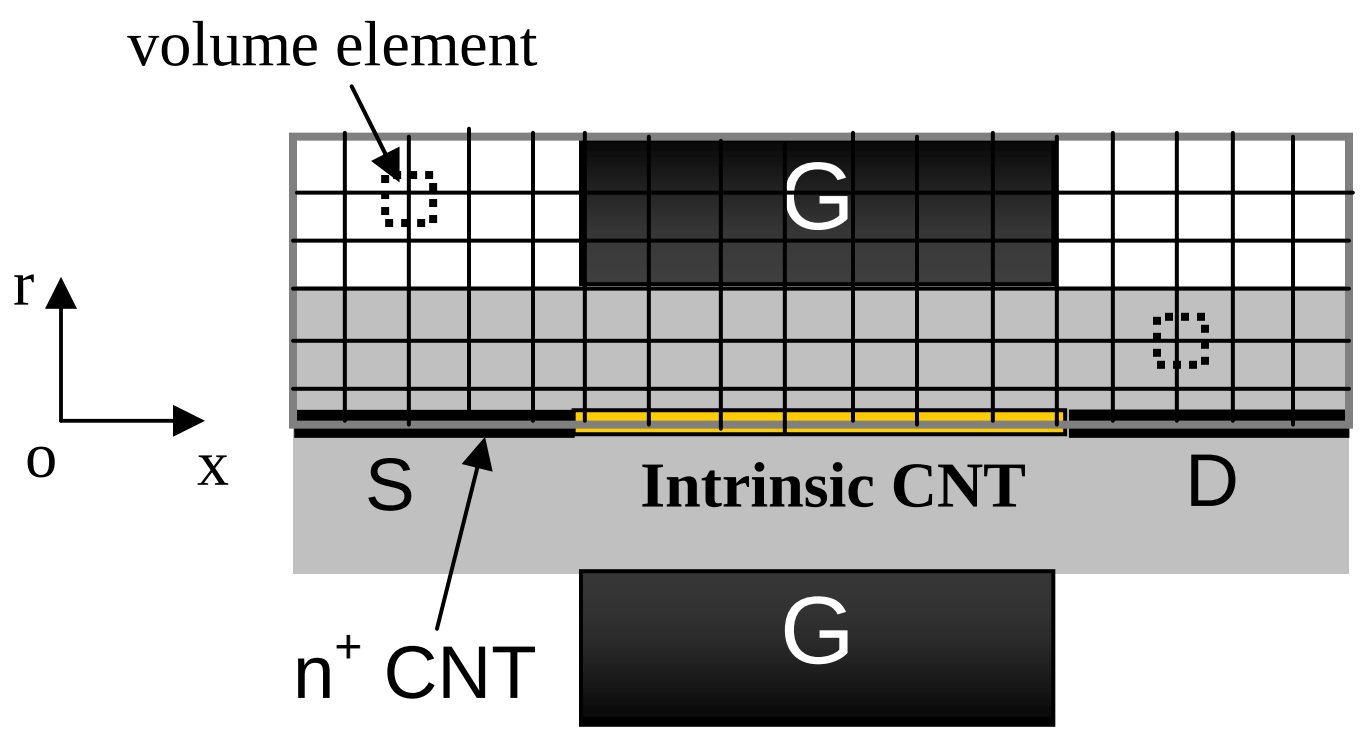

Fig. 6 The modeled, coaxially gated carbon nanotube transistor with heavily-doped, semi-infinite nanotubes as the source/drain contacts. The channel is intrinsic and the gate length equals the channel length. Also shown are the simulated area, the simulation grid and the cylindrical coordinate system used for solving the Poisson equation. The dashed rectangular area shows the volume element used to discretize the Poisson equation at ($x_i$, $r_j$).

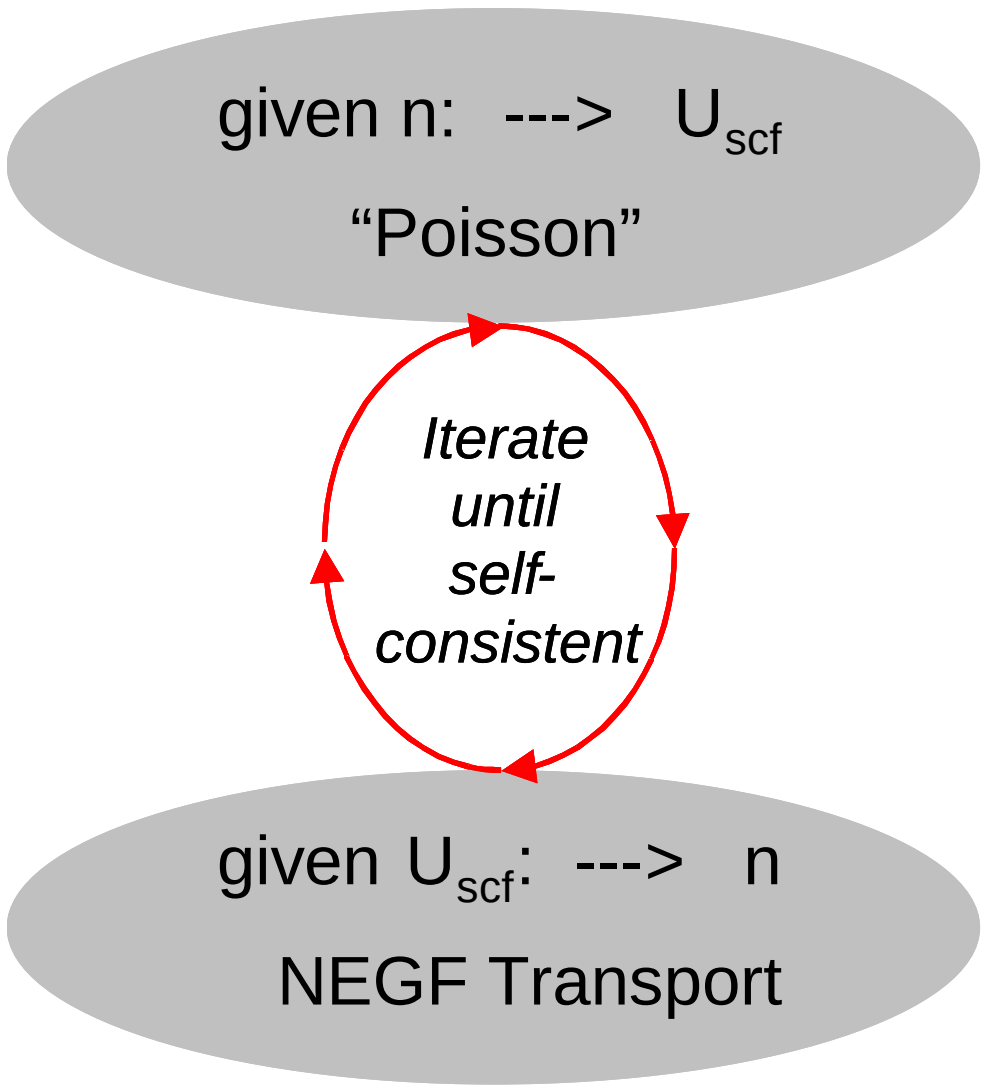

Fig. 7 The self-consistent iteration between the NEGF transport and the electrostatic Poisson equation. These two equations are iteratively solved until self-consistency is achieved. Then the current is computed using the self-consistent potential.

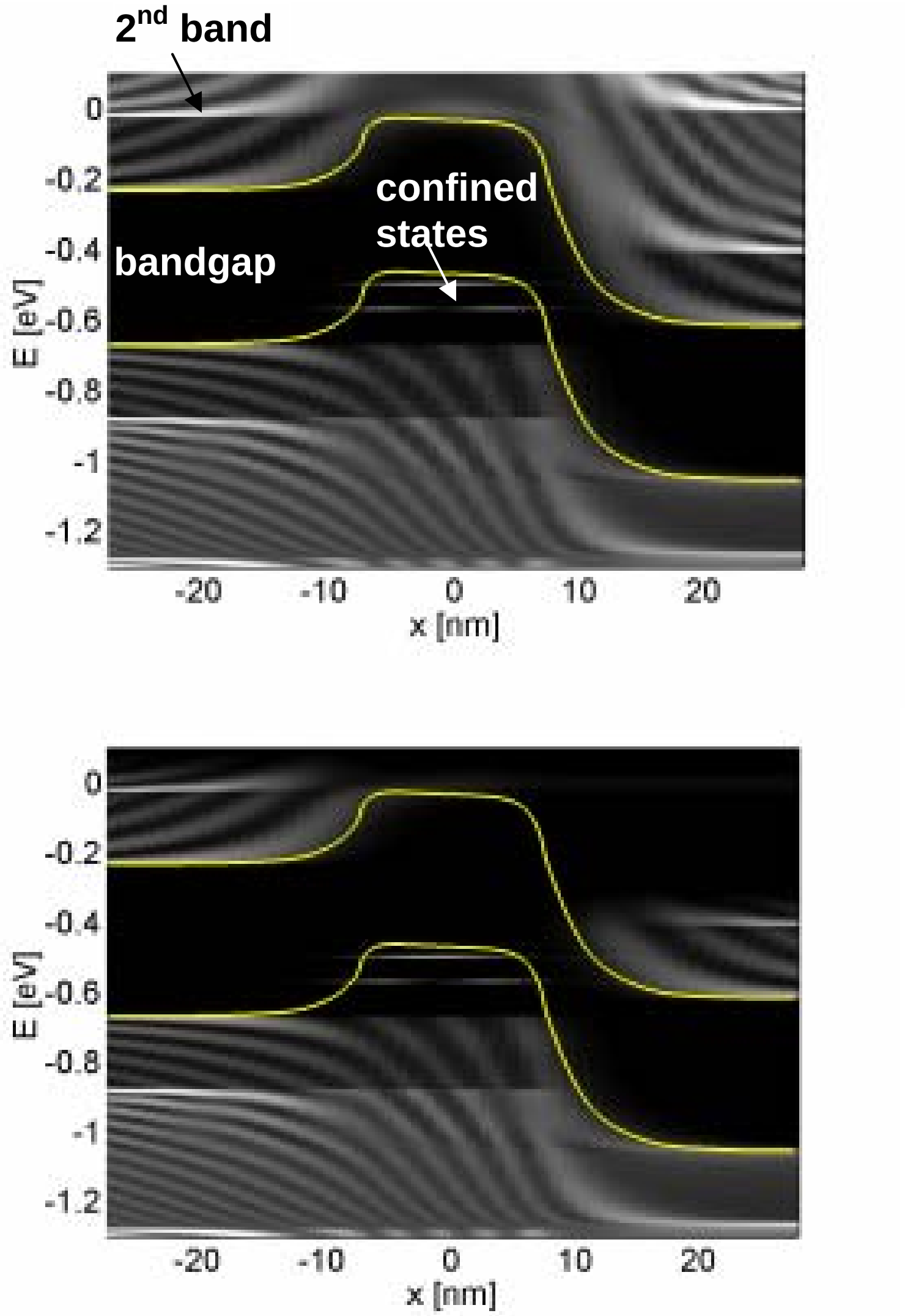

Fig. 8 (a) The local-density-of-states (LDOS) and (b) the electron density spectrum computed by the real space approach at $V_G$=0.25V and $V_D$=0.4V. The modeled transistor is shown in Fig. 7a. The nanotube is a (25,0) CNT with a diameter d~2nm and bandgap Eg~0.4eV.

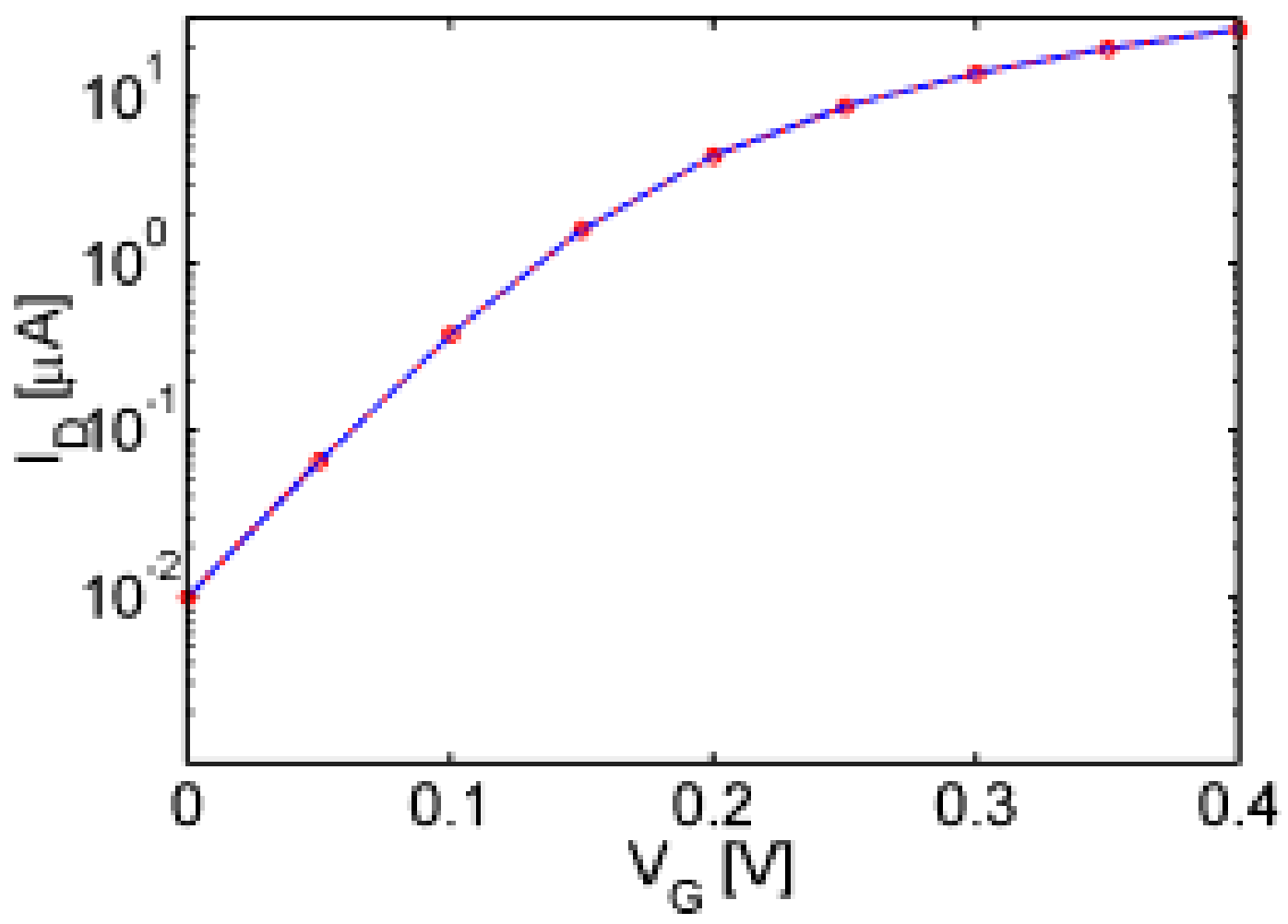

Fig. 9 The I-V characteristics computed by the real space approach (the solid line) and the mode space approach with 2 subbands (the circles) for a CNTFET as shown in Fig. 7. The (13,0) nanotube channel length is 15nm.

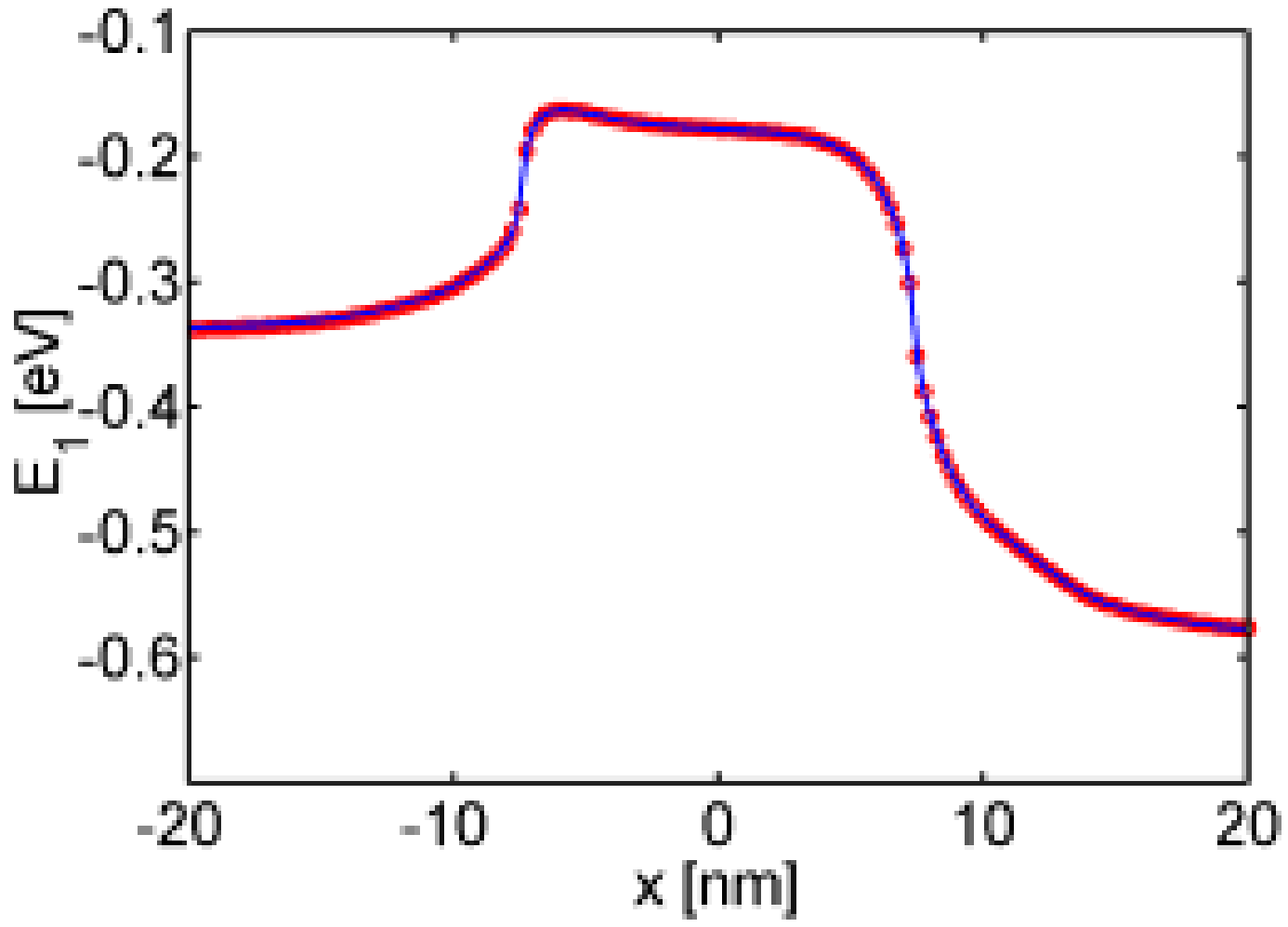

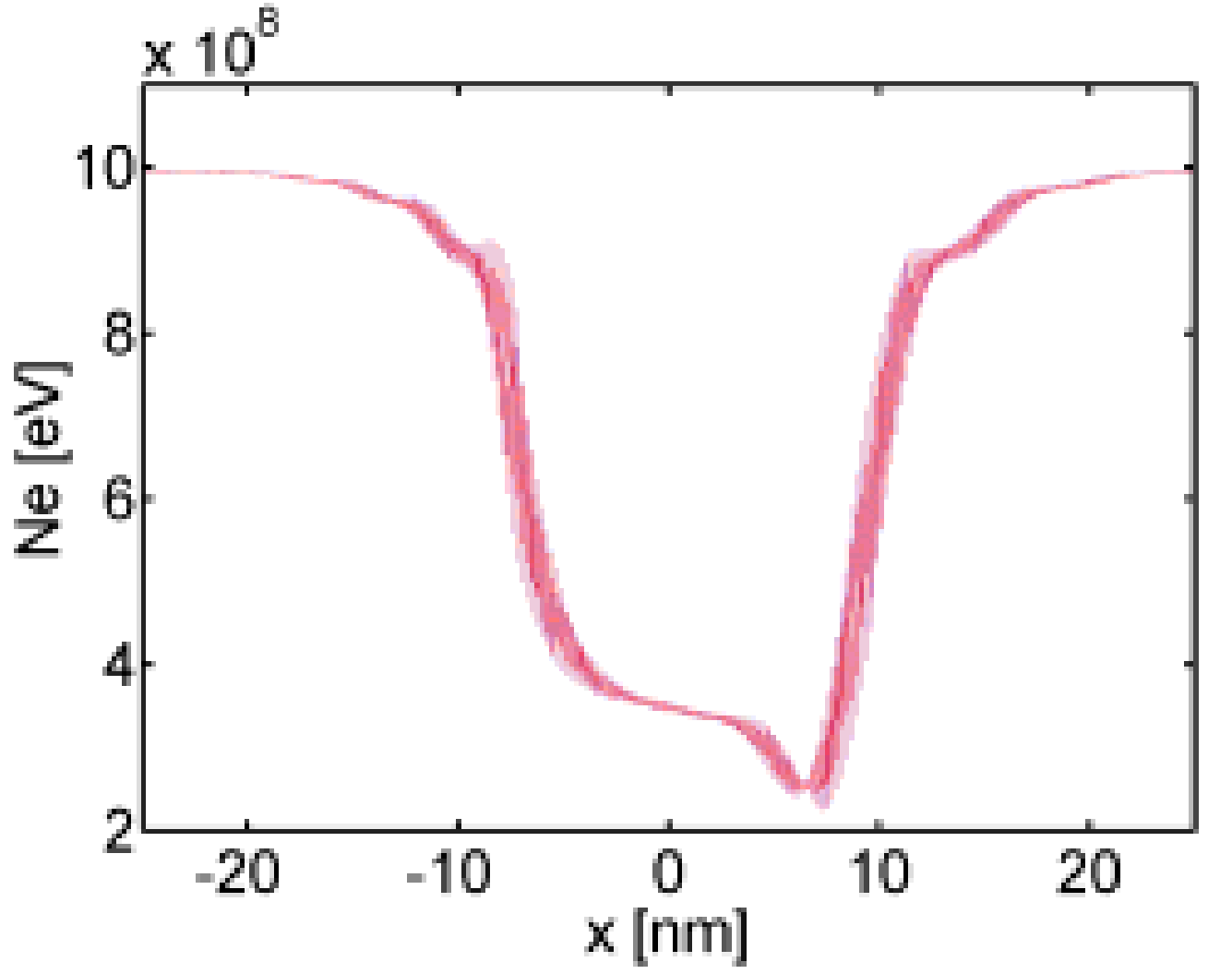

Fig. 10(a) The conduction band profile computed by the real space approach (the solid lines) and the mode space approach (the circles) at $V_G = V_D = 0.4\,\text{V}$. (b) The charge density computed by the real space approach (the solid line) and the mode space approach (the dashed line). The solid and dashed lines lie on top of each other.

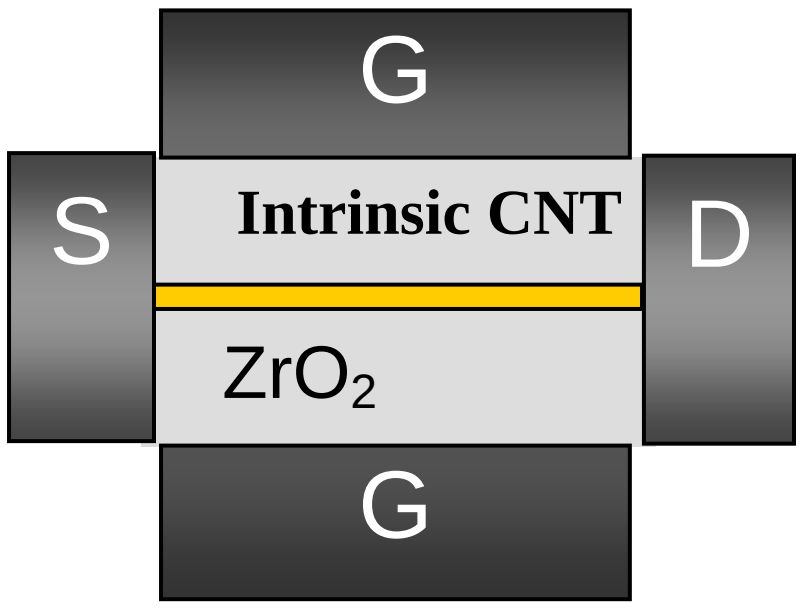

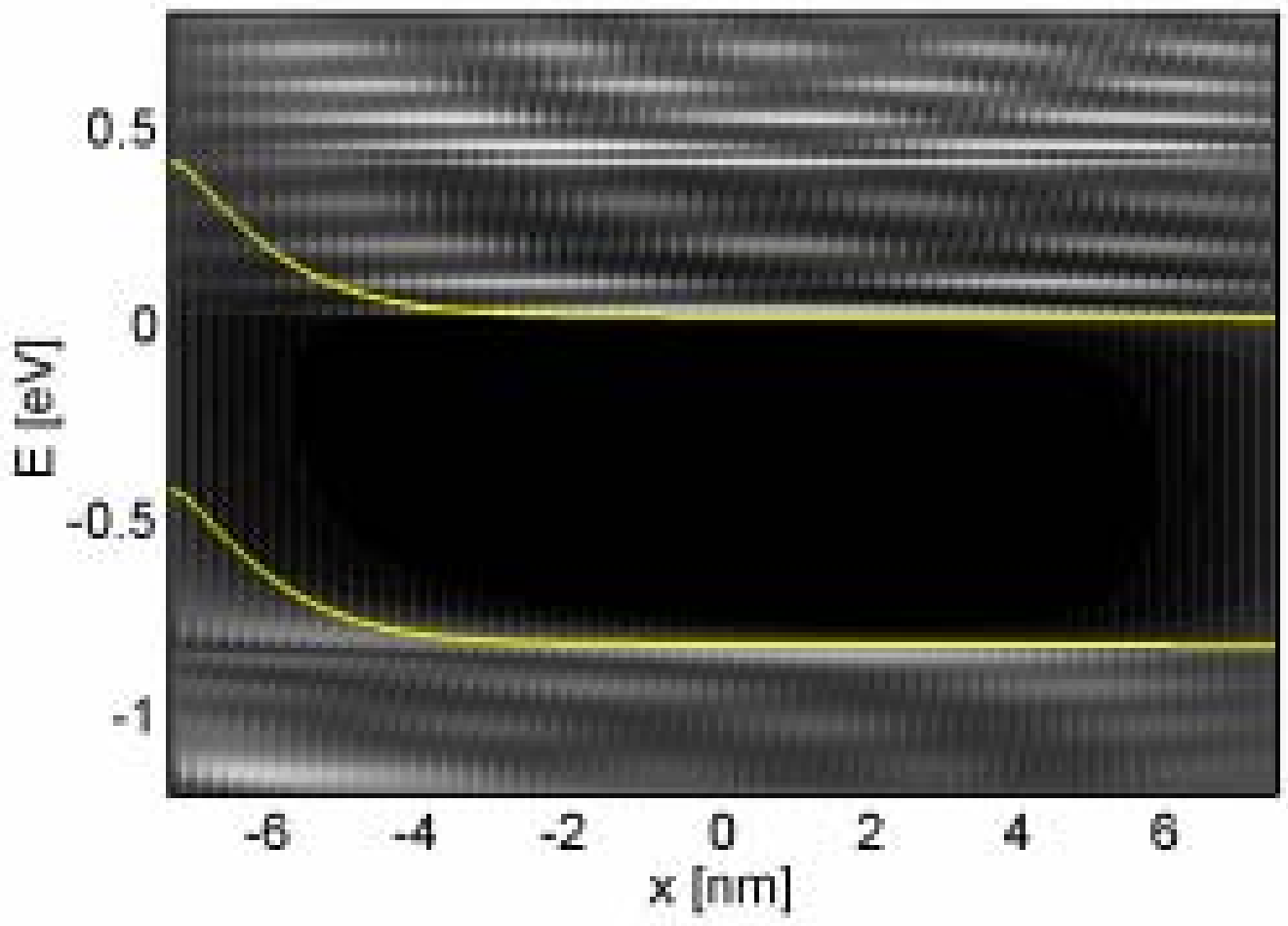

Fig. 11(a) The coaxially gated Schottky barrier carbon nanotube transistor with an intrinsic nanotube channel directly attached to metal source and drain contacts. The nanotube channel is a (13,0) zigzag CNT with a diameter d~1nm and band gap Eg~0.83eV. The gate insulator is a 2nm-thick $ZrO_2$. (b) The local-density-of-states (LDOS) at $V_D = V_G = 0.4$ V, which clearly shows tunneling through the Schottky barrier at the source end of the channel, and metal induced gap states (MIGS) at the metal/CNT interfaces.

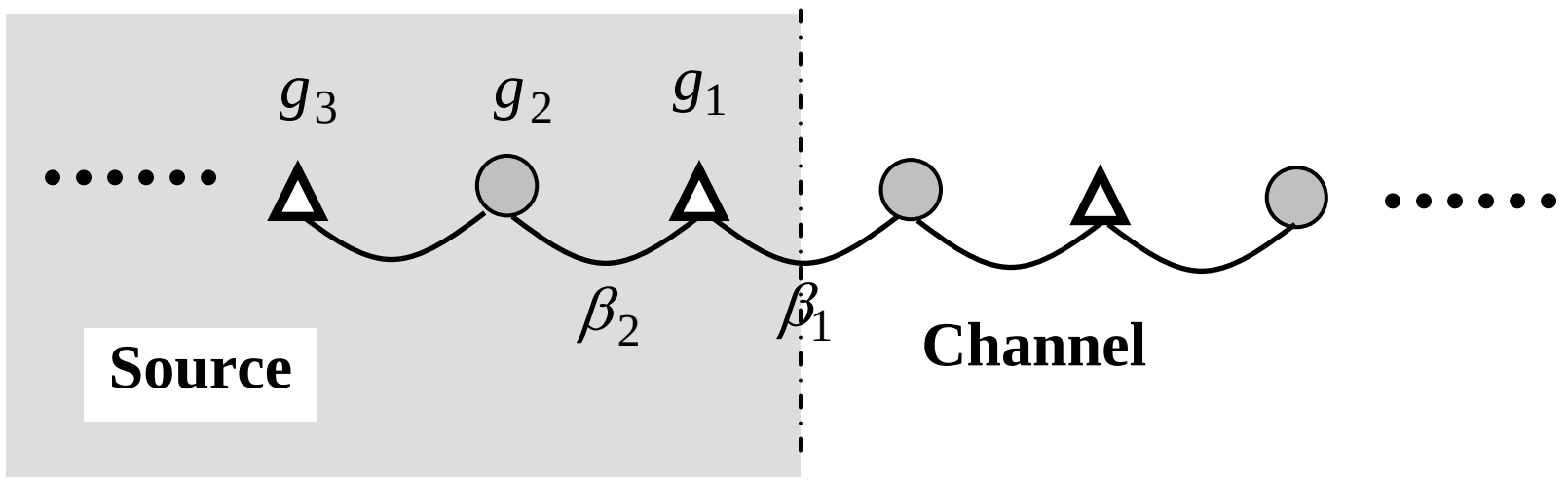

Fig. A1 Computing the source self-energy for a zigzag nanotube. The circles represent A-type carbon rings and the triangles represent B-type carbon rings. $g_i$ is the surface Green's function for the *i*th carbon ring inside the source. $\beta_1$ ( $\beta_2$ ) is the first (second) kind coupling matrix between neighboring rings, as described in the text.